\documentclass[a4paper,fleqn]{cas-sc}

\usepackage[authoryear]{natbib}
\usepackage{rotating}
\usepackage{lineno}
\usepackage{setspace}
\usepackage{appendix}
\usepackage{tcolorbox} 
\usepackage{mdframed}

\def\tsc#1{\csdef{#1}{\textsc{\lowercase{#1}}\xspace}}
\tsc{WGM}
\tsc{QE}
\tsc{EP}
\tsc{PMS}
\tsc{BEC}
\tsc{DE}

\begin{document}
\let\WriteBookmarks\relax
\def\floatpagepagefraction{1}
\def\textpagefraction{.001}

\shorttitle{K.I. Dale et al. 2024}

\shortauthors{K.I.Dale et al.}

\title [mode = title]{Compositional Outcomes of Earth Formation from a Narrow Ring}                      
\tnotemark[1]

\tnotetext[1]{This project was funded by the ERC HolyEarth, grant number: 101019380}

%
\author[1]{Katherine I. Dale}[type=editor,
                        auid=000,bioid=1,
                        orcid=0000-0001-7511-2910]

\cormark[1]

\ead{katherine.dale@oca.eu}


\credit{Conceptualisation of this study,
Methodology, code, manuscript and analysis of results}

\affiliation[1]{organization={Université Côte d'Azur, Observatoire de la Côte d'Azur, CNRS, Laboratoire Lagrange},
    city={Nice},
    postcode={06302}, 
    country={France}}

\author[1,2]{Alessandro Morbidelli}
\credit{Conceptualisation of this study,
methodology and code }
\author[3]{David C. Rubie}
\credit{Methodology and code}

\author[4]{David Nesvorný}
\credit{Data and editorial}

\affiliation[2]{organization={Collège de France, CNRS, PSL Univ., Sorbonne Univ.},
    city={Paris},
    postcode={75014}, 
    country={France}}
    
\affiliation[3]{organization={Bayerisches Geoinstitut, University of Bayreuth},
    city={Bayreuth},
    postcode={95490}, 
    country={Germany}}

\affiliation[4]{organization={Solar System Science \& Exploration Division, Southwest Research Institute},
    city={Boulder},
    postcode={CO 80302}, 
    country={USA}}
\cortext[cor1]{Corresponding author}

\begin{abstract}
The origin of Earth's formation material remains controversial. Here we address the problem from the elemental point of view. We use an approach similar to that of \cite{Rubie2015}, with the technical improvements presented in \cite{Dale2023} to simulate the chemical evolution of the Earth's mantle during a series of metal-silicate partial equilibration events associated with accretional collisions. However, we introduce two radical differences. First, we consider the dynamical model in which Earth forms from a dense ring of planetesimals and planetary embryos near 1 AU, with a low-density extension of the planetesimal population into the asteroid belt. Second, we divide the ring and asteroid belt population into four zones. The zone closest to the Sun is assumed to be populated by planetesimals and embryos enriched in elements more refractory than Si relative to CI concentrations and fully depleted in volatile elements including S and C. This material is not sampled in the meteorite record, except potentially in angrites. Moving further away from the Sun, the remaining three zones are populated by material with the compositions of enstatite chondrites, ordinary chondrites and CI chondrites respectively. Using this model, we fit the chemical composition of the bulk silicate Earth in terms of relative abundances of the oxides of Al, Mg, Fe, Si, Ni, Co, Nb, V, Cr, W, Mo and C by adjusting the boundaries of the above compositional zones and the refractory enrichment in the innermost zone thus giving us four compositional free parameters. A fifth and final fitting parameter concerns the depth of planetesimal equilibration in a magma ocean produced following a giant impact. We considered twenty-two simulations of the ring model, all of which produced at least one planet of similar mass and semi-major axis as the Earth. Each simulation of the chemical evolution of the Earth assumed either an initially hot or cold target for collisions producing a total of forty-eight Earth analogues. Seventeen of the simulations resulted in a planet with bulk mantle chemistry quite similar to that of the observed bulk silicate Earth (BSE) despite their differing hierarchical growth sequences. However, these differences in the sequence, size and initial position of the impactors result in different values of the five fitting parameters. This equates to differences in the proportion of each meteorite type required to obtain a chemical composition similar to the BSE, though some similarities remain such as the requirement for the Earth to accrete the majority (60-80\%) of its material from the innermost refractory enriched region. This implies that, paired with the right hierarchical growth sequence and with some constraints, more than one ring model compositional structure can produce an Earth analogue consistent with the BSE, with very similar final mantle compositions for all considered elements. 
\end{abstract}

\begin{highlights}
\item The formation of the terrestrial planets from a ring of planetesimals and embryos, with a low mass extension towards the asteroid belt, is viable from the point of view of Earth's chemical composition 
\item The chemistry of the bulk silicate Earth (BSE) reveals the need for a compositional component which is unsampled in the meteorite collection, that is very reduced and refractory enriched 
\item The bulk composition of the Earth is not fully constrained by the BSE composition but also depends on the conditions and properties of metal silicate equilibration events
\end{highlights}

\begin{keywords}
earth formation \sep mantle composition \sep accretion \sep metal-silicate differentiation \sep early solar-system
\end{keywords}

\maketitle
\section{Introduction}

The source of the material that formed Earth remains a topic of significant debate. The Earth is often assumed to be constructed from material similar to that contained in the meteorite collection and thus we should be able to form the composition of the bulk silicate Earth (BSE) using known meteoritic compositions. However, when considering the classic scenario of terrestrial planet formation, in which the planets grow by accretional collisions of planetesimals and planetary embryos, the supra-chondritic Al/Si and Mg/Si ratios in the BSE suggest that Earth accreted a substantial fraction of its material from bodies enriched in elements more refractory than Si with respect to CI meteorites and to enstatite (EC) and ordinary chondrites (OC) \citep{Dauphas2015, Morbidelli2020}. 

In this work, we aim to address the formation of the Earth from an elemental point of view, creating an Earth analogue from a mixture of known meteorite groups (CI, EC, OC) as well as a “missing component” represented by refractory-enriched, material depleted in volatiles including S and C. The angrite parent body could be representative of this "missing component" \citep{Tissot2022} while Marrocchi et al., (in review) identifies it as PO chondrules. In broad terms, this should be material that condensed and/or was reprocessed in the disc at high temperature. We employ a model adapted from \cite{Rubie2015}, which links Earth's geochemical evolution to its accretion history as determined by dynamical simulations, by simulating metal-silicate equilibration following each accretionary impact. Our model extends this by using the improvements described in \cite{Dale2023}. Specifically, we use the results of smooth particle hydrodynamic (SPH) simulations of \cite{Nakajima2021} to calculate both the mass and the basal pressure of the magma ocean generated by each giant collision between planetary embryos. Planetesimals and embryo impacts are treated differently, with the material carried by planetesimals being stored near the surface of the target embryo until a giant impact occurs. This impact forms a magma pond, where part of the impacting embryo’s core equilibrates with part of the target's molten silicate, which then relaxes into a global surface magma ocean in which the metal previously delivered by planetesimals (both to the impacting and target embryos) can finally equilibrate with the silicate. 

\section{Methods}
\subsection{Ring Set-Up}

\begin{figure}[htbp!]
    \centering
    \includegraphics[width=1\linewidth]{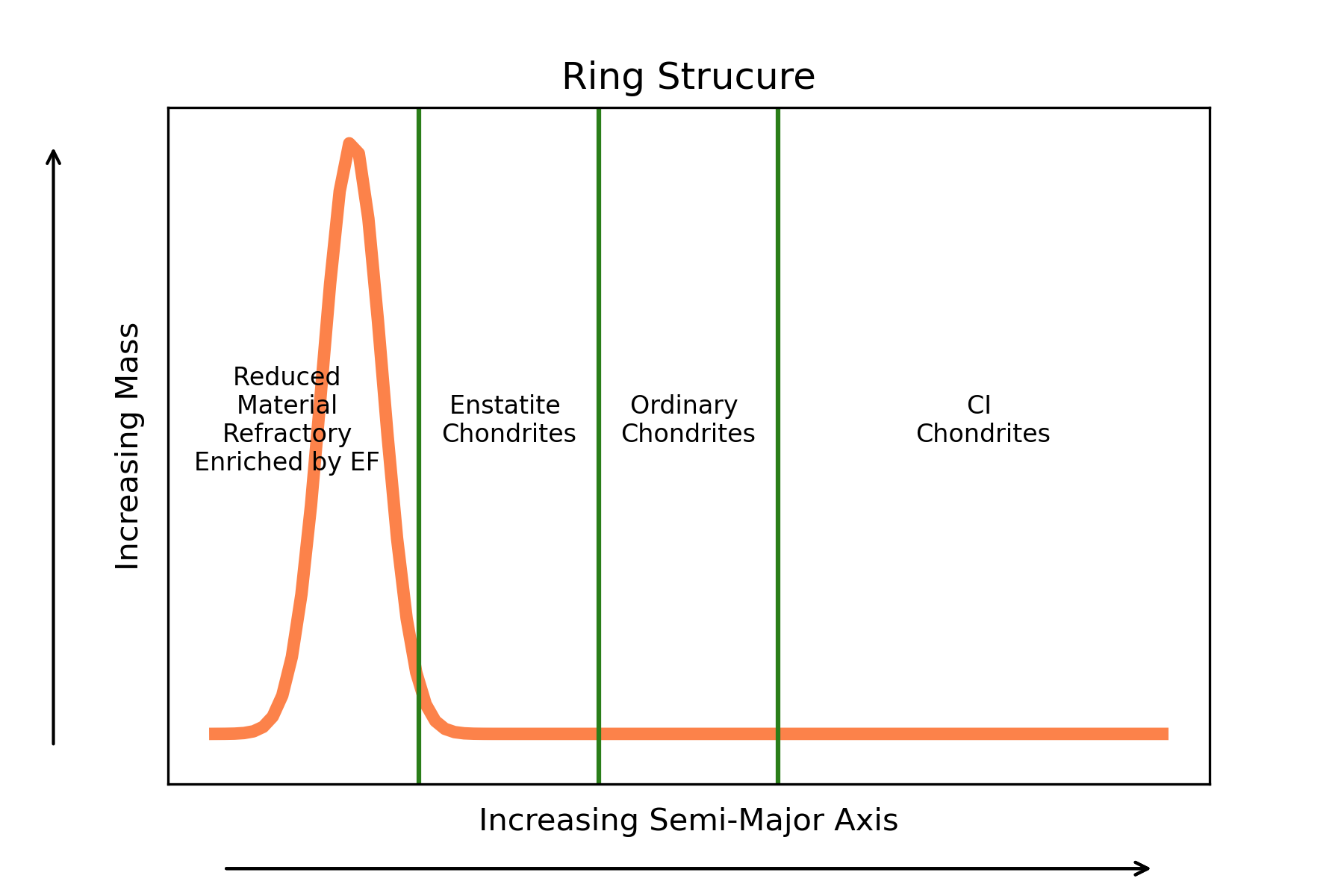}
    \caption{A cartoon illustration of the structure of the narrow ring in which the terrestrial planets form. Most of the mass is concentrated between 0.7 and 1 AU with a low mass extension towards the asteroid belt \citep{Nesvorny2021}. We split the ring and the extension into four distinct compositional regions which are, in order of increasing semi-major axis: material enriched in elements more refractory than Si by a factor of EF, enstatite chondrites, ordinary chondrites and CI chondrites. This cartoon does not represent our final results and is only to show an example of how a ring in our simulations may have looked.}
    \label{fig:Ring_Structure}
\end{figure}

\begin{table*}[hbt!]
    \centering
  \caption{The five free fitting parameters allowed to vary from simulation to simulation.}
  \label{tab:FreeParameters}
  \begin{tabular}{| >{\raggedright}p{2cm} | >{\raggedright\arraybackslash}p{2.4cm} | >{\raggedright\arraybackslash}p{2.4cm} | >{\raggedright\arraybackslash}p{2.4cm} | >{\raggedright\arraybackslash}p{2.4cm} | >{\raggedright\arraybackslash}p{2.4cm} |}
    \hline
     Free Parameter & Distance 1 (D1) & Distance 2 (D2) & Distance 3 (D3) & Enrichment Factor (EF) & Planetesimal Equilibration Factor (PEF) \\ \hline
     Explanation & Boundary between the regions of refractory enriched material and the enstatite chondrite material & Boundary between the regions of enstatite chondrite material and ordinary chondrite material  & Boundary between the regions of ordinary chondrite material and CI chondrite material  & Enrichment in elements more refractory than Si of innermost material  & Fraction of the pressure at the base of the global magma ocean following a giant impact at which planetesimals equilibrate \\ \hline
     Range & 0.3 - 4 AU (D1<D2)  & 0.3 - 4 AU (D2<D3) & 0.3 - 4 AU  & >1  & 0.15 - 1\\ \hline
    \end{tabular}
\end{table*}

\begin{figure}[htbp!]
    \centering
  \includegraphics[width=0.65\textwidth]{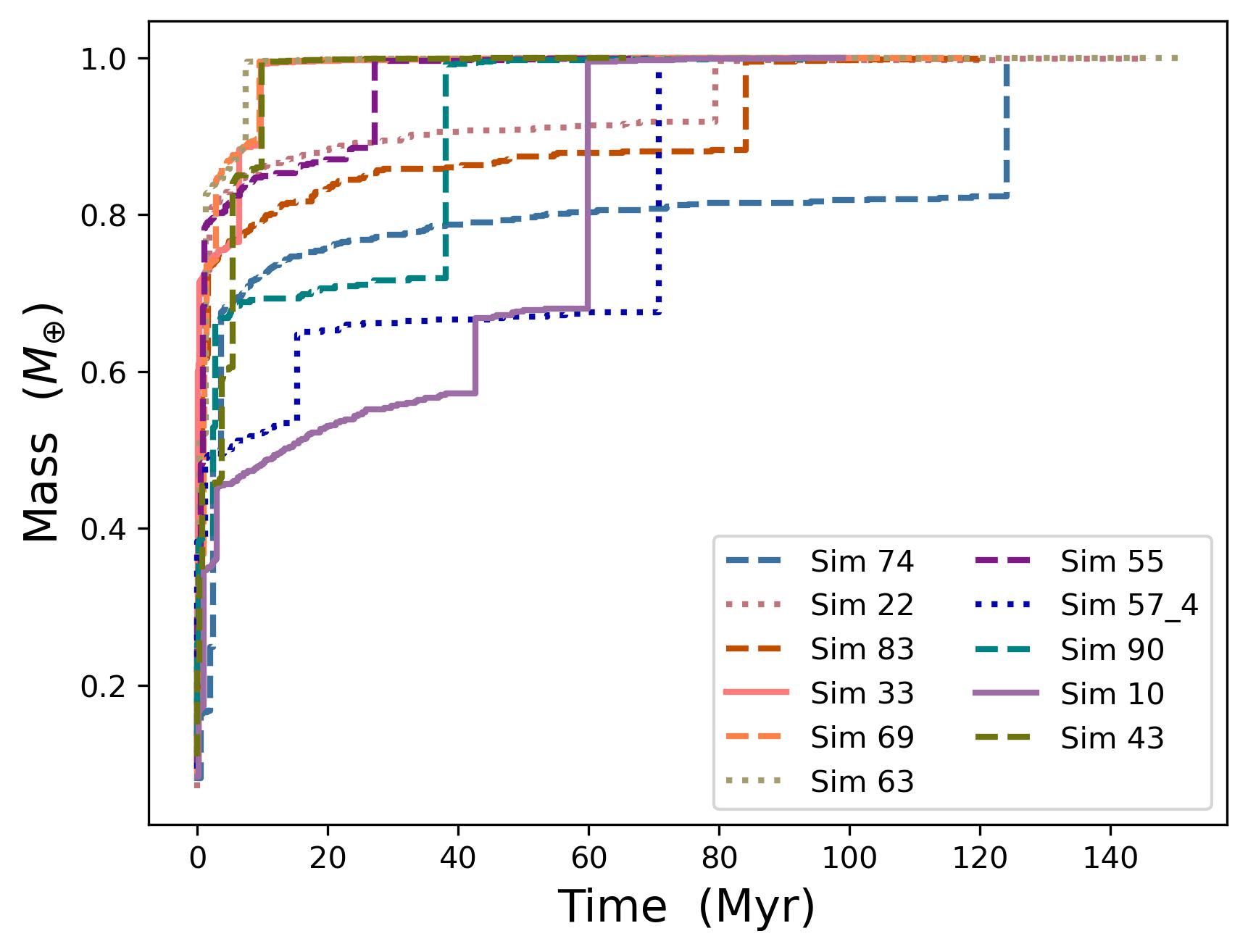} 
  \caption{The growth of Earth analogues in good fit simulations (see \autoref{Results}). The nature of the giant collisions (hot or cold target) does not affect planet growth.}
  \label{fig:Earth_Growth}
\end{figure}

The model of metal-silicate equilibration takes the results of N-body accretion scenarios that simulate the growth of the terrestrial planets. In previous works (\citealp{Rubie2015, Rubie2025,Jennings2021,Dale2023} etc.) these N-body formation scenarios used the Grand Tack model \citep{Walsh2011}. In this work, we aimed to examine the ring model scenario for the formation of terrestrial planets. This paradigm has been explored in \citet{Hansen2009}, \citet{Izidoro2022}, \citet{Nesvorny2021} and \citet{Woo2023,Woo2024} among others. These simulations start from a narrow, dense ring of planetary embryos and planetesimals located around 1 AU. Embryos are assumed to exist only between 0.7 AU and 1 AU, thus concentrating the mass from which the terrestrial planets originate (see \autoref{fig:Ring_Structure} for a cartoon illustration of the material may exist in the ring). We examine simulations from \cite{Nesvorny2021}, which feature this narrow ring of planetesimals and embryos and a low mass extension of planetesimals into the asteroid belt. Each simulation starts with 20 embryos as well as 1000 planetesimals, many of which are lost to the Sun or the giant planet region throughout the simulation. The embryos are allowed to collide with each other and with planetesimals as in the standard model of hierarchical growth of the terrestrial planets (e.g., \citealp{ChamberandWetherill1998}). The simulations start at the end of the gas-disc phase (approx. 10 My), with the giant planets assumed to be fully formed, on initially circular orbits which then rapidly evolve to the current orbital configuration via a phase of dynamical instability. In contrast to many Grand Tack simulations, the concentration of planetesimals and embryos results in a system where Earth-mass objects form typically in less than 100 Myr. Giant collisions (between two planetary embryos) are also less violent on average when compared to Grand Tack simulations. Each embryo has an initial mass of just under a tenth of Earth's mass while the initial planetesimal mass is just under 0.2\% of Earth's mass, though each planetesimal should be interpreted as representing a large number of smaller bodies throughout our reanalysis of these simulations.  The masses of planetesimals and embryos are then scaled for each simulation so that the final mass of the Earth analogue is exactly 1 Earth mass. This rescaling normally represents a change in mass of just a few percent. This small change is unlikely to affect the outcome of the N-body simulations.  Moreover, planetesimals that impact the Earth-analogue planet after the final giant impact have their mass scaled so that the mass of this late accretion is around 0.5\% of the final Earth mass. This is to match the late veneer mass inferred from the abundance of highly siderophile elements in the BSE \citep{Chou1978,Jacobson2014,Fischer_Goode2012}. We started with a suite of 100 N-body simulations. 22 of these simulations produced a terrestrial planet population similar to that of the Solar System with one Venus analogue, one Earth analogue and at least one Mars analogue. The Earth analogue planet was then chosen based on its mass and the occurrence on this planetary body of the latest giant impact in the simulation, representing the Moon-forming impact. 20 of the simulations produced only one planet that fit the criteria to be considered an Earth-analogue and 2 simulations formed 2 planets that fit the criteria. This gives a total of 24 Earth-analogue planets, the growth of a selection of these is shown in \autoref{fig:Earth_Growth}.

\begin{table*}[hbt!]
  \centering
  \caption{The compositions of bodies in each distinct region of the ring with meteoritic data from \cite{Alexander2019A, Alexander2019B}. Carbon contents are assigned following \cite{Blanchard2022} with undifferentiated NC bodies containing 80 times more carbon than their differentiated counterparts. Undifferentiated CC bodies contain 1500 times more carbon than differentiated CCs. In the table, all bodies are embryos and thus differentiated, though CCs contain no metal and have lost carbon due to heating. Refractory enriched bodies are entirely carbon-depleted. Water contents for CIs are taken from (\cite{Alexander2019A}) while OCs contain a small quantity of water (e.g. \cite{Hutchison1987}, \cite{Piani2015}). Quantities marked with a * are enriched relative to CI ratios by a factor of 1.3 in this example but this parameter is free to vary through the model. Mantle oxides are listed in the first half of the table and are highlighted in grey. Core metals are in the lower half. Initial partitioning occurs at 0.65 x core mantle boundary pressure and the associated temperature though we find this makes little difference to the initial compositions of the bodies. BSE oxide abundances are from \citet{Palme&ONeill2003} with the exception of Mo which is from \citet{Greber2015}}.
  \label{tab:Compostiosn}
  \renewcommand{\arraystretch}{0.9} 

    \begin{tabular}{| p{1.5cm} | p{1.8cm} | p{1.8cm} | p{1.8cm} | p{1.8cm} | p{3.15cm} |}
    \hline
    & \multicolumn{4}{|c|}{Zone Compositions (wt\%)} & \\ \hline
Component & Refractory \newline Enriched & Enstatite Chondrite & Ordinary Chondrite & CI Chondrite & BSE \\ \hline
\rowcolor{LightGrey}
    Al  &  3.548228*  & 1.897557  & 2.170483   & 1.710768 & 4.49 $\pm$ 0.359 \\ \hline
\rowcolor{LightGrey}
    Mg   & 26.201399   & 22.481138    & 23.782349   & 17.054456  & 36.77 $\pm$ 0.368\\ \hline
\rowcolor{LightGrey}
    Ca  & 2.849528*  & 1.481915  & 1.770680  & 1.373892 & 3.65 $\pm$ 0.292 \\ \hline
\rowcolor{LightGrey}
    Na  & 0.000  & 0.935748  & 0.892294  & 0.720640  & 0.259 $\pm$ 0.013\\ \hline
\rowcolor{LightGrey}
    Fe  & 0.000  & 0.000  & 21.133129  & 25.652474 & 8.1 $\pm$ 0.050 \\ \hline
\rowcolor{LightGrey}
    Si  & 31.459577  & 39.881963  & 37.691678  & 24.671080 & 45.4 $\pm$ 0.318 \\ \hline
\rowcolor{LightGrey}
    Ni  & 0.022336  & 0.021299  & 0.015769  & 1.453873 & 0.186 $\pm$ $9.3 \times 10^{-3}$ \\ \hline
\rowcolor{LightGrey}
    Co  & 0.001093  & 0.001054  & 0.000780  & 0.071128  & 0.0102 $\pm$ $5.1 \times 10^{-4}$\\ \hline
\rowcolor{LightGrey}
    Nb  & $1 \times 10^{-6}$* & 0.000  & $1 \times 10^{-6}$  & $4.4 \times 10^{-5}$  & $5.95\times10^{-5}$ $\pm$ $1.19\times10^{-5}$ \\ \hline
\rowcolor{LightGrey}
    V & $1.8 \times 10^{-4}$*  & $9 \times 10^{-5}$  & $1.08 \times 10^{-4}$  & $8.658 \times 10^{-3}$  & $8.6\times10^{-3}$ $\pm$ $4.3\times10^{-4}$\\ \hline
\rowcolor{LightGrey}
    Cr  & $5.673 \times 10^{-3}$   & $4.543 \times 10^{-3}$  & $4.786 \times 10^{-3}$  & 0.369284  & 0.252 $\pm$ 0.0252\\ \hline
\rowcolor{LightGrey}
    Pt  & $2 \times 10^{-6}$*  & $1 \times 10^{-6}$  & $1 \times 10^{-6}$  & $1.01 \times 10^{-4}$  & $7.6\times10^{-7}$ $\pm$ $1.52\times10^{-7}$\\ \hline
\rowcolor{LightGrey}
    Pd  & $1 \times 10^{-6}$ & $1 \times 10^{-6}$  &$1 \times 10^{-6}$  & $6.9 \times 10^{-5}$ & $7.1\times10^{-7}$ $\pm$ $1.42\times10^{-7}$ \\ \hline
\rowcolor{LightGrey}
    Ru & $2 \times 10^{-6}$* & $1 \times 10^{-6}$  &$1 \times 10^{-6}$  & $6.9 \times 10^{-5}$ & $7.4\times10^{-7}$ $\pm$ $1.48\times10^{-7}$ \\ \hline
\rowcolor{LightGrey}
    Ir & $1 \times 10^{-6}$* & $1 \times 10^{-6}$  &$1 \times 10^{-6}$  & $5.4 \times 10^{-5}$ & $3.5\times10^{-7}$ $\pm$ $3.5\times10^{-8}$ \\ \hline
\rowcolor{LightGrey}
    W & 0.000* & 0.000  & 0.000 & $1.3 \times 10^{-5}$  & $1.2\times10^{-6}$ $\pm$ $5.04\times10^{-7}$\\ \hline
\rowcolor{LightGrey}
    Mo & $3 \times 10^{-6}$* & $2 \times 10^{-6}$  &$1 \times 10^{-6}$  & $1.38 \times 10^{-4}$ & $2.3\times10^{-6}$ $\pm$ $1.38\times10^{-6}$  \\ \hline
\rowcolor{LightGrey}
    H$_2$O & 0.000 & 0.000  & 0.973456 & 21.556782 & 0.1 $\pm$ 0.03\\ \hline
\rowcolor{LightGrey}
    S & 0.000 & 0.000  & 0.000 & 11.520290 & 0.02 $\pm$ $5.0 \times10^{-3}$ \\ \hline
\rowcolor{LightGrey}
    C & 0.000 & 0.000  & 0.000 & 0.005027  & 0.014 $\pm$ $4.0 \times10^{-3}$  \\ \hline

    Fe & 30.634604 & 27.986974  & 5.475691 & 0.000 & \\ \hline

    Si & 3.012127 & 0.837689  & 0.000 & 0.000 &\\ \hline

    Ni & 1.737727 & 1.656994  & 1.226818 & 0.000 & \\ \hline

    Co & 0.085083 & 0.082035  & 0.060714 & 0.000 & \\ \hline

    Nb & $6.3 \times 10^{-5}$* & $2.9 \times 10^{-5}$  &$3.6 \times 10^{-5}$  & 0.000  & \\ \hline

    V & 0.012084* & 0.006030  & 0.007257  & 0.000  & \\ \hline

    Cr & 0.429514 & 0.343894  & 0.362359  & 0.000 & \\ \hline

    Pt & $1.91 \times 10^{-4}$* & $1.34\times 10^{-4}$  &$1.12 \times 10^{-4}$  & 0.000  &\\ \hline

    Pd & $9.2 \times 10^{-5}$ & $8.8 \times 10^{-5}$  &$6.5 \times 10^{-5}$  & 0.000 & \\ \hline

    Ru & $1.53 \times 10^{-4}$* & $9.4 \times 10^{-5}$  &$7.9 \times 10^{-5}$  & 0.000 &\\ \hline

    Ir & $1.02 \times 10^{-4}$* & $5.6 \times 10^{-5}$  &$5.1 \times 10^{-5}$  & 0.000 & \\ \hline

    W & $2.1 \times 10^{-5}$* & $1.5 \times 10^{-5}$  &$1.4 \times 10^{-5}$  & 0.000  &\\ \hline

    Mo & $2.13 \times 10^{-4}$* & $1.15 \times 10^{-4}$  & $1.11 \times 10^{-4}$  & 0.000 & \\ \hline

    S & 0.000 & 2.378344  & 4.429226  & 0.000  &\\ \hline

    C & 0.000 & 0.002195  & 0.001947  & 0.000 & \\ \hline
\end{tabular}
\end{table*}

\begin{table*}[hbt!]
    \centering
  \caption{The oxidation of each distinct region of the ring. Iron oxidation is taken from the Urey-Craig diagram (e.g. \citealp{Krot2014}). Silicon oxidation in enstatite chondrites is calculated using \citet{Keil1968} and \citet{Lehner2014}.}
  \label{tab:Oxidation}
  \begin{tabular}{|c|c|c|c|c|}
    \hline
    & \multicolumn{4}{|c|}{Zone Oxidation (\%)} \\ \hline
     & Refractory Enriched & Enstatite Chondrite & Ordinary Chondrite & CI Chondrite \\ \hline
     Percentage Si in Metal & 17  & 4.3  & 0.0  & 0.0 \\ \hline
     Percentage Fe in Metal & 100  & 100  & 25  & 0.0 \\ \hline
    \end{tabular}
\end{table*}

As stated in the introduction, Earth is believed to have accreted a substantial fraction of its material from bodies enriched in elements more refractory than Si with respect to CI meteorites, which are representative of the composition of the Sun. Thus, we split the ring of embryos and planetesimals and its extension toward the asteroid belt into four distinct compositional regions, in order of increasing semi-major axis: a region enriched in elements more refractory than Si, an enstatite chondrite region, an ordinary chondrite region and a CI chondrite region. An idealised cartoon representation of these regions and the boundaries separating them is shown in \autoref{fig:Ring_Structure}. The locations of these boundaries, the refractory enrichment of the innermost region (EF) and a value referred to as the planetesimal equilibration factor (PEF, see below) are five free parameters in the model that we tune to find an Earth-analogue with a mantle closest to the observed composition of the BSE. We minimise the difference between the known BSE values of the oxides of Al, Mg, Fe, Si, Ni, Co, Nb, V, Cr, W, Mo and C and the abundances of these constituents in our Earth analogue mantles. Thus, the number of constraints (12) is much larger than the number of free parameters (5). Notice also that we do not consider volatile elements (e.g. Na, Zn, ...) in our fit. The consistency of our results with the Earth's volatile budget will be qualitatively discussed in \autoref{Results}. By varying the three distances the quantity of material in each region changes and thus the percentage of material from each region that is accreted to an Earth-analogue changes. The distances are varied independently of each other with the constraint that D1 < D2 < D3 so that regions can not overlap.

 For the innermost region, we assign a CI composition, to which we increase the relative abundance of elements more refractory than Si by a factor EF and remove more volatile elements (particularly water, S and C). The rationale for this choice is that the bodies in this region should have formed in the inner, hot part of the disc, where volatiles could not condense. Because the enrichment in refractory elements should be interpreted as the consequence of an incomplete condensation of Si and other elements of equivalent condensation temperature (e.g. Fe. Mg. ...) the same enrichment factor EF is applied to all refractory elements. Notice that the suppression of the volatile elements increases the bulk abundance of all other elements. For instance, in the example of \autoref{tab:Compostiosn}, Si comprises 17.7 wt\% and Fe 30.6 wt\% of the bodies of the innermost zone, while they only comprise 11.5 wt\% and 20 wt\% in CI meteorites, respectively. We use data from \citet{Alexander2019A, Alexander2019B} to determine the elemental compositions of the three sampled meteorite types (\autoref{tab:Compostiosn}). In addition to the elemental composition of each region, we require the oxidation of each distinct meteoritic region. This is defined by the fraction of total iron and silicon in metal in each of these regions; a value that was fitted for as a gradient in previous versions of the model (\citealp{Rubie2015,Dale2023} etc.). The values of these are straightforward for enstatite, ordinary and carbonaceous chondrites (and are listed in \autoref{tab:Oxidation}). These fractions for the unsampled region must be estimated. The fraction of silicon initially in metal, in the region closest to the Sun, is a value fitted for by \citet{Rubie2015} and other works using similar models. The resulting values which produce the best Earth analogues normally lie between 14\% and 20\%. We have taken an average of these values as 17\% and this is used to define the fraction of total silicon in metal in the refractory enriched region.

The fraction of total iron in metal is trickier to estimate. We consider that this innermost region was highly reduced, similar to the enstatite chondrites. The fraction of silicon in metal for this region is also consistent with a highly reduced region. Thus we assume that, in this very reduced region, all iron is in metal. The imposed values of the fraction of iron in metal and the fraction of silicate in metal for each compositional zone of the ring are fixed and listed in \autoref{tab:Oxidation}.

\begin{figure}[hbtp!]
    \centering
    \includegraphics[width=0.8\linewidth]{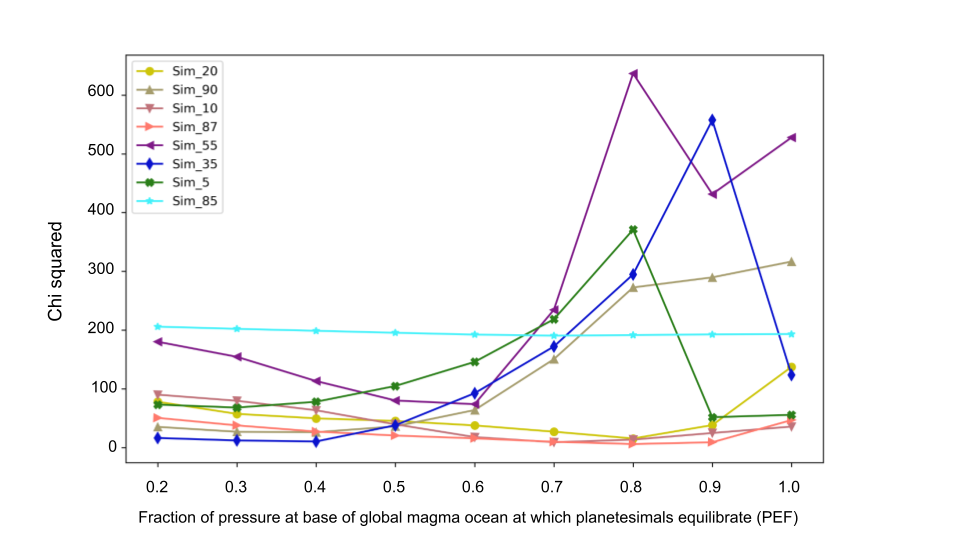}
    \caption{Dependence of the quality of the BSE fit (expressed as $\chi^2$) on the PEF which regulates the depth in the global magma ocean at which the metal delivered by planetesimals equilibrates.}
    \label{fig:Planetesimal_Depths}
\end{figure}

\subsection{Metal-silicate Equilibration Model}
The model used follows that of \cite{Rubie2015}, which coupled, for the first time, the chemical evolution of the Earth to the accretion history of the simulation that formed the Earth analogue. \cite{Dale2023} then expanded this approach, making improvements on the treatment of embryo-embryo impacts and metal silicate fractionation mechanisms following planetesimal impacts.

\begin{figure}[hbtp!]
\centering
  \includegraphics[width=0.8\textwidth]{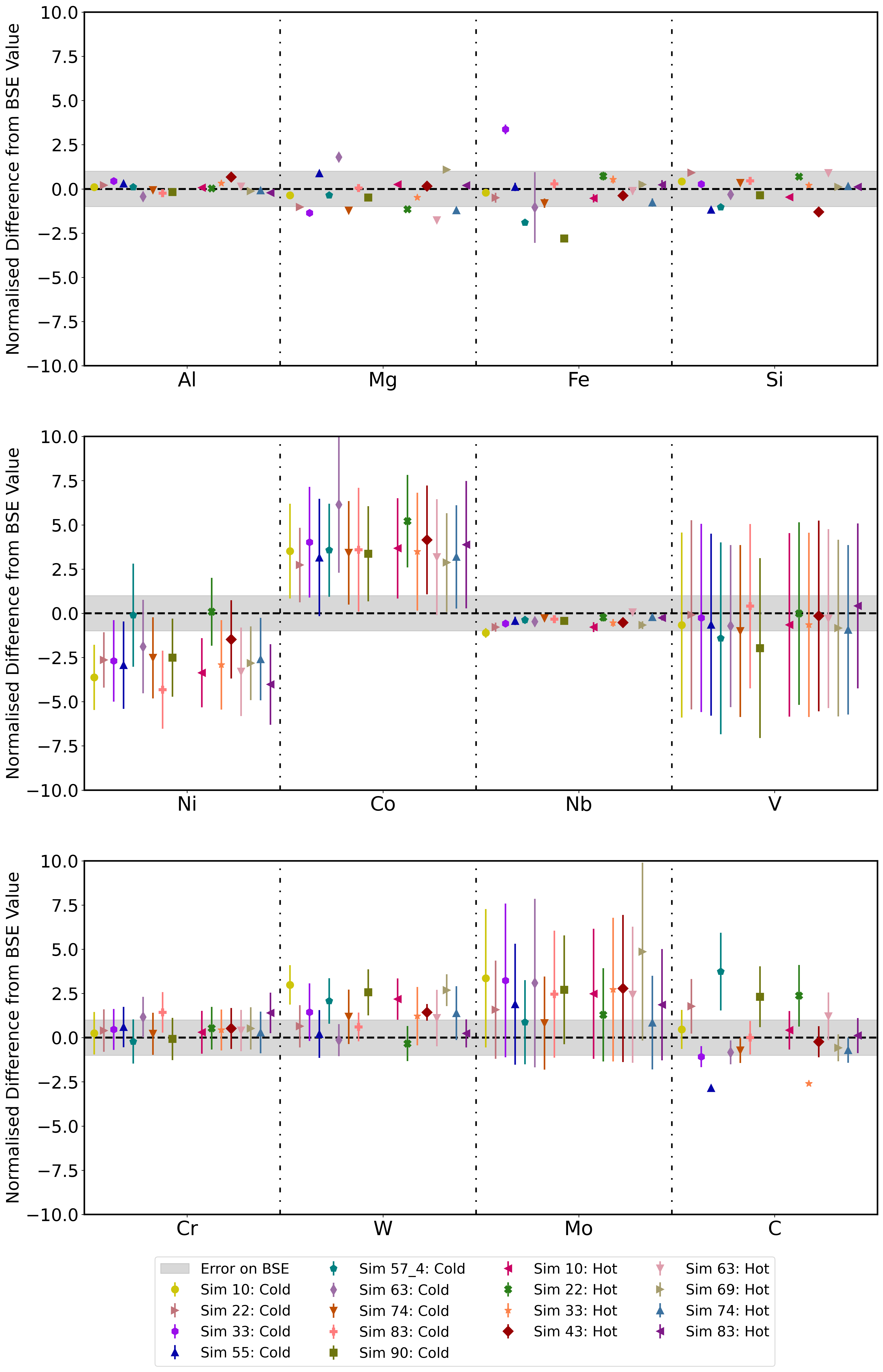} 
  \caption{The compositions of the Earth-analogues produced in simulations that give a good fit. Uncertainties on the model are shown with error bars while uncertainties on the observed value for Earth are shown as a grey band. In all panels, the differences between the model values and the BSE values have been normalised by the uncertainty on the BSE value using \autoref{Plot Equation}. Thus, if the normalised differences are less than 1 (grey band) then the model is consistent with the data.}
  \label{fig:Elements}
\end{figure}

In the terrestrial planet accretion model there are two types of collisions: embryo-embryo collisions are highly energetic and cause large-scale melting of the target and thus the development of a deep magma pond. Planetesimal collisions are smaller and far less energetic and are thus less likely to cause significant melting when hitting a target embryo. Thus, their materials are assumed to be stored near the surface of their target until an embryo-embryo collision occurs. The assumption of a solid target is reasonable given that giant impacts are distributed further apart than the typical timescales for magma ocean crystallisation of less than 5 Myr \citep{ElkinsTanton2008}. In our model, embryo-embryo collisions are treated as described in \citet{Dale2023}, with the equilibration of the impactor's metal with part of the target's silicate at the bottom of the impact-induced magma pond. The amount of mantle melt produced, and the pressure at the base of the magma pond are calculated using the scaling laws of \citet{Nakajima2021}. This pressure is associated with a temperature that lies approximately midway between the solidus and liquidus temperatures of peridotite at that pressure \citep{Rubie2015}. On impact, the projectile embryo core separates from its mantle and sinks through the magma pond. It equilibrates with a portion of the molten target silicate given by \citet{Deguen2014} at the pressures and temperatures given by the previously mentioned scaling laws. The oxygen fugacities are calculated based on mass balance as in  \citet{Rubie2011}. The impactor metal that is not oxidised during the metal-silicate equilibration reaction is then added to the core. The silicates that are involved in the equilibration reaction from the projectile are added to the target silicate. The magma pond then hydrostatically relaxes into a global magma ocean in which the previously delivered planetesimal metal equilibrates with the silicates. Following \citet{Dale2023} and \citet{Rubie2025}, metal from the planetesimals is equilibrated by the "dispersed" metal-silicate fractionation mechanism (described in detail by \citet{Rubie2025} and first proposed as "model 2" in \citet{Rubie2003}), which is crucial for successfully modelling final mantle concentrations of W and Mo. The metal should be dispersed as small droplets in the magma ocean. Metal-silicate equilibration is assumed to occur progressively at a mechanical boundary layer at the base of the magma ocean because vertical convection velocities are close to zero \citep{MartinNokes1988}. However, because the crystallisation timescale of the magma ocean is short (probably within 5 Myr) a significant fraction of the metal delivered by planetesimals between giant impacts may equilibrate at pressures and temperatures lower than those corresponding to the initial magma ocean depth. While equilibration still occurs progressively following the fractionation model just described, the characteristic pressure and temperature of the mechanical boundary layer will change with magma ocean crystallisation. In the absence of a reliable model of crystallisation timescales, we define an effective equilibration pressure for planetesimal material which we call the planetesimal equilibration factor (PEF) so that equilibration occurs at a pressure that is PEF multiplied by the pressure at the base of the initial global magma ocean. A value of PEF close to 1 means that planetesimal metal equilibrates immediately after the formation of the surface magma ocean when the latter has its maximal depth determined from \citet{Nakajima2021}. A smaller value indicates that planetesimal metal equilibrates when a magma ocean has partially crystalised with values close to 0 indicating that planetesimal metal equilibrates, on average, close to the end of magma ocean crystallisation. We varied this fraction across eight randomly chosen simulations to see what value gives the best fit for the BSE (the lowest $\chi^2$ value) in the majority of these simulations. The results are presented in \autoref{fig:Planetesimal_Depths}. They show that the best-fit PEF changes significantly from one simulation to another. Four simulations (Sim\_5, Sim\_87, Sim\_10 and Sim\_20) have their best BSE composition occurring when the planetesimals are equilibrated at between 70\% and 100\% of the pressure at the base of the initial magma ocean. Three simulations (Sim\_55, Sim\_90 and Sim\_35) produce the best Earth when planetesimals are equilibrated at between 20\% and 40\% of the pressure at the base of the initial magma ocean. The composition of the Earth produced by Sim\_85 appears to not be dramatically affected by the pressure at which planetesimals equilibrate. Given these results, we have incorporated PEF as a fifth fitting parameter for our model, allowing it to vary between simulations. 
 
\begin{figure}[htbp!]
    \centering
  \includegraphics[scale=0.485]{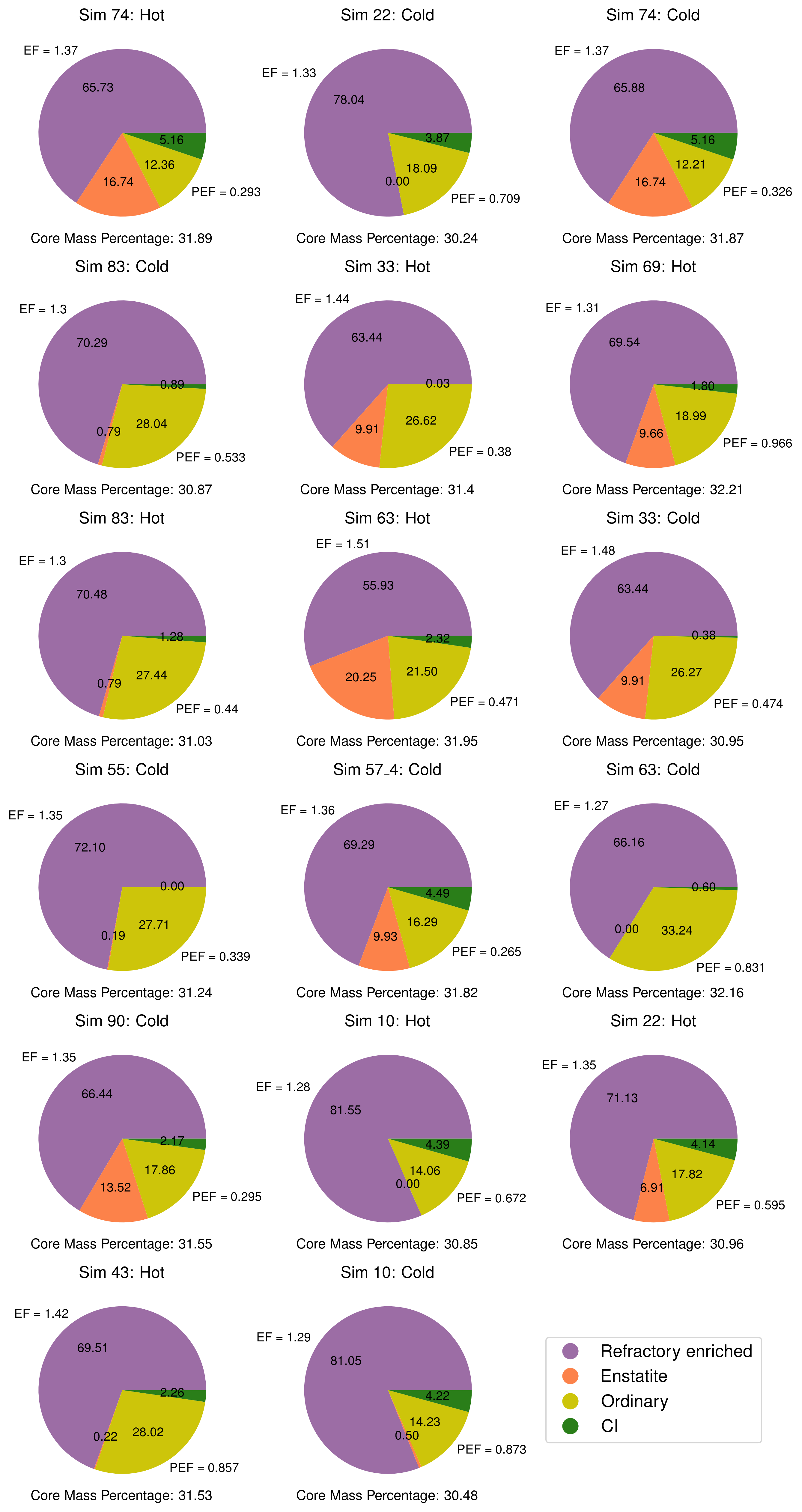} 
  \caption{The different contributions from the four compositional zones which accrete to form each good fit Earth analogue planet. Refractory enrichment for the unsampled material (EF) is shown next to the refractory enriched region of each pie chart. PEF for each simulation is also indicated to the right of each chart. Below each chart is the percentage of the final planet mass (1 Earth mass) that is contained within the analogue core.}
  \label{fig:PieCharts}
\end{figure}

\section{Results and Dicussion} \label{Results}
We tested each of the 24 Earth analogues produced by dynamical simulations. Because the properties of the magma pond/ocean generated during giant impacts depend on the thermal state of the target before the collision \citep{Nakajima2021}, for each Earth analogue we do two chemical evolution simulations: one assuming that the target is always hot (2000 K at the surface) and one assuming it is always cold (300 K at the surface). Thus, in total, we model 48 Earths. We identified our good-fit simulations based not only on the $\chi^2$ value of the fit to the BSE composition but also by examining how many of the 12 fitted (Al, Mg, Fe, Si, Ni, Co, Nb, V, Cr, W, Mo and C) oxides have concentrations overlapping with BSE values when we account for uncertainties. In \autoref{fig:Elements}, the concentrations obtained in our model are renormalised as:

\begin{equation} \label{Plot Equation}
  C^{model}_{norm} =  \frac{C^{model}- C^{BSE}}{\Delta C^{BSE}}
\end{equation}

where $\Delta C^{BSE}$ is the uncertainty on BSE measurements of $C^{BSE}$ (listed in \autoref{tab:Compostiosn}). The grey band in \autoref{fig:Elements} thus represents $|C^{model}-C^{BSE}| <= \Delta C^{BSE}$.
 The uncertainties on the BSE come from the measurements and are listed in \autoref{tab:Compostiosn}. The uncertainties in the model results come solely from the propagation of the uncertainties on the partition coefficients determined in laboratory experiments. Thus, they correspond to the intrinsic uncertainties of the model, which cannot be reduced. We do not consider the model uncertainties due to poor knowledge of the assumed fixed parameters (i.e. those not determined by best fit). A simulation that fits 10 of these 12 components is classed as a 'good fit'. In total, 17 of the 48 simulations tested produced a good fit. Their compositions are shown in \autoref{fig:Elements}. We do not fit for Ca as it is a refractory lithophile element and acts similarly to Al. If aluminium oxide can be fit in a model then calcium oxide can be fit as well. Considering both would artificially increase the number of constraints and reduce the resulting $\chi^2$ of the fit while not providing extra information as to the goodness of fit. Furthermore, H$_2$O is not included in our fit given that we do not account for its loss in atmospheric formation and thus most simulations produce mantles with too much water systematically.

\begin{figure}[htbp!]
    \centering
  \includegraphics[scale=0.3]{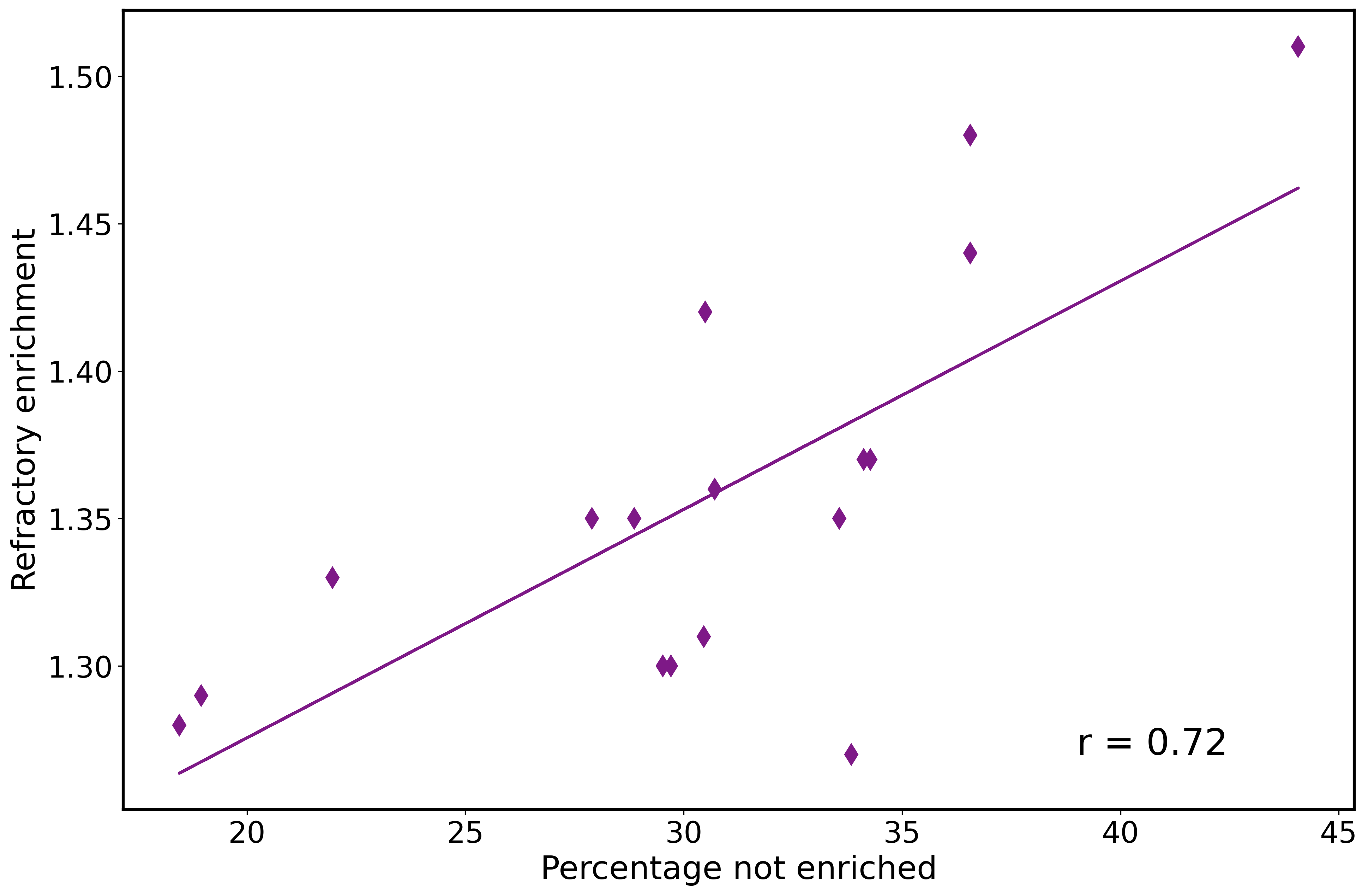} 
  \caption{The amount of refractory enrichment in the innermost compositional region of the ring (EF) compared with the fraction of the bulk Earth-analogue's composition accreted from the other regions, i.e. material that is not enriched. Each point represents an Earth analogue that successfully reproduced the BSE. The r-value indicating the significance of the correlation is listed.}
  \label{fig:Enrichment_Changes}
\end{figure}

There are a few key features of \autoref{fig:Elements} that should be noted. For most simulations, fits for oxides of Al, Mg, Fe, Si, Nb, V and Cr are good. Mo concentrations are too high in all simulations though acceptable given the large errors on the model results. W can be fit well in over half of the simulations. Both of these results are significant improvements with respect to the results of \citet{Dale2023} and \citet{Jennings2021} who found Mo and W very difficult to fit. These improvements are related to using distinct meteoritic compositions. In previous model versions, with the exception of \citealp{Rubie2025}, all material was assumed to have the composition of the refractory enriched material with only the oxidation and volatile content changing with semi-major axis. This updated model uses four distinct compositions, with four distinct concentrations of Mo and W. This gives the model more freedom to adjust the compositional boundaries to change the amount of Mo and W accreted by the Earth. Concentrations of Ni and Co mirror each other such that when one is high the other is low. This makes it difficult to fit both simultaneously, with less than half of the good fit simulations in \autoref{fig:Elements} achieving an acceptable fit (i.e. with overlapping error bars) for both Ni and Co. Interestingly, most of the successful simulations require a low PEF, suggesting low equilibration pressure may be the key to achieving the concentrations of Ni and Co in the BSE. We get a good value for carbon compared to the BSE in most of the simulations. However, two simulations (Sim\_55 Cold and Sim\_33 Hot) have almost zero concentrations of carbon in their Earth-analogue mantles. This result is likely to be related to the near zero percentages of CI chondrites accreted to these Earth-analogues (see \autoref{fig:PieCharts}) and would likely increase significantly with the addition of 1-2 planetesimals from the CI region, without affecting the goodness of the fit. 

The compositions of the bodies that accrete to form our 17 good fit Earth analogues are shown in \autoref{fig:PieCharts}. There are some universal features in the makeup of the good fits: they all have the majority of Earth material derived from the refractory enriched, volatile depleted region and the second highest contribution from the ordinary chondrite region. However, the pie charts do show three distinct groups which are as follows: 1) an Earth accreted almost entirely from ordinary (\textasciitilde 30\%) and refractory enriched material (\textasciitilde 70\%) (Sim\_83 Cold, Sim\_83 Hot, Sim\_55 Cold, Sim\_63 Cold and Sim\_43 Hot), 2) an Earth made mostly of ordinary and refractory enriched material (\textasciitilde 80\%) but with the addition of around 4\% CI material (Sim\_22 Cold, Sim\_10 Cold and Sim\_10 Hot), 3) an Earth mostly made of ordinary and refractory enriched material with a significant percentage of enstatite material (8-20\%) and a varying percentage of CI material (Sim\_74 Hot, Sim\_74 Cold, Sim\_33 Hot, Sim\_69 Hot, Sim\_63 Hot, Sim\_33 Cold, Sim\_57\_4 Hot, Sim\_90 Cold and Sim\_22 Hot). Despite these differences in the relative contributions of the considered four sources of material (and therefore in the bulk Earth composition), there are no significant differences in the final model BSE compositions, because they all fit the actual BSE comparatively well. This suggests that the differences between bulk compositions of the 17 different good fit simulations are compensated by different degrees of metal-silicate equilibration defined by the properties of the sequence of collisions that make up the Earth analogue in each simulation. We stress that none of the pies in \autoref{fig:PieCharts} depicts a unique reservoir for Earth material even though the fitting procedure of the compositional zone boundaries would have allowed this to occur. This is because none of the four compositional zones can reproduce the BSE alone: the meteorite groups have different bulk compositions and the enriched zone lacks water, S and C, regardless of the assumed EF value. Even if we had defined a zone with a material identical to that of the bulk Earth, the BSE would not be reproduced by that zone alone. The reason is explained in \citet{Rubie2011}: Earth needs to start by accreting reduced material and then accrete more oxidised compositions. Forming Earth from a material with the oxygen fugacity of the current Earth would thus not result in the BSE composition. This means that Earth's accretion had to comprise different reservoirs of material. This includes the need for a highly reduced zone, enriched in elements more refractory than Si, which makes up a large portion of all our analogue BSEs. It may be tempting to believe that we could recreate the Earth using only the compositions found in the meteorite reservoir (enstatite chondrites, ordinary chondrites and CI chondrites). However, when this is tested we find very poor fits in comparison to those shown in \autoref{fig:Elements} (see \autoref{fig:NoEnrich}). This highlights the necessity for this fourth reservoir of material during Earth's accretion when examining the ring model.

We observe a relationship between the required enrichment (EF) of the refractory enriched region and the best-fit compositional pie. \autoref{fig:Enrichment_Changes} shows how EF differs with the composition of the material that accretes to a good fit Earth analogue, with an r-value calculated to show the strength of the correlation.  There is a strong positive correlation when we compare the sum of all non-enriched materials (enstatite, ordinary and CI) to EF (\autoref{fig:Enrichment_Changes}). This seems obvious given the refractory enrichment of the target BSE is the same throughout all simulations. The fact that the correlation here is not stronger suggests that factors affecting metal-silicate equilibration also affect the required enrichment. 

\begin{figure}[htbp!]
    \centering
  \includegraphics[width=0.985\textwidth]{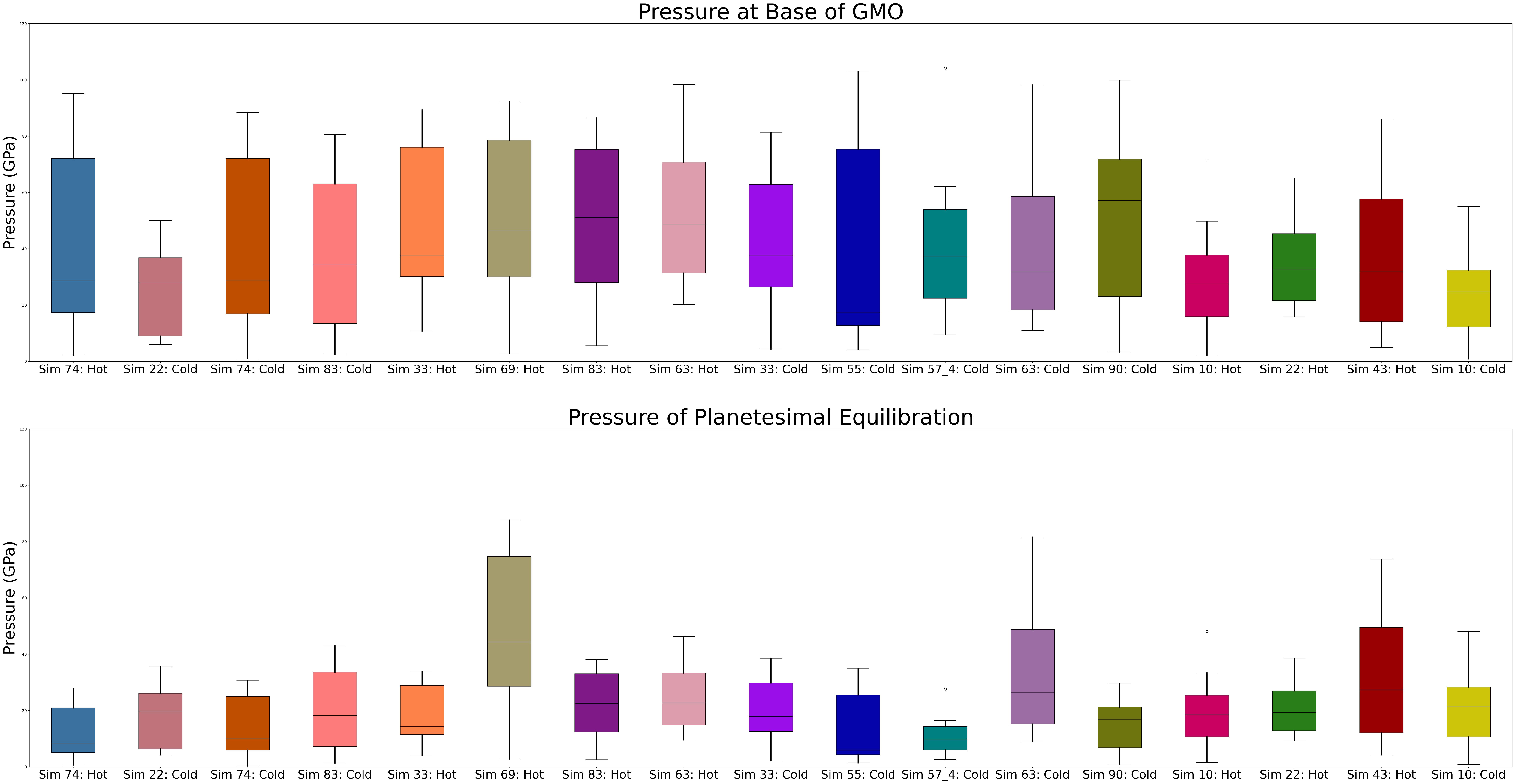} 
  \caption{Top: The range of pressures at the base of the hydrostatically relaxed magma ocean after each giant impact. Bottom: The pressures at which planetesimal material actually equilibrates, calculated using a fraction, PEF, of the above values. In all box plots, the central line shows the median values, 50\% of all data (interquartile range) is contained within the box and the whiskers show the lowest and highest values which are within 1.5 times the interquartile range. Points represent outliers. }
  \label{fig:PEFS}
\end{figure}

The value of PEF, the fraction of the pressure at the base of the hydrostatically relaxed magma ocean at which planetesimal metal equilibrates, for each good fit simulation is listed next to the appropriate pie chart (\autoref{fig:PieCharts}). These vary greatly from simulation to simulation, ranging from 0.265 to 0.951. \autoref{fig:PEFS} shows the pressures at the base of the global magma ocean through each simulation and the range of pressures at which planetesimal material equilibrates (equal to the former value multiplied by PEF). This figure shows that different PEF values come from the need to have a median equilibration pressure for planetesimals between 10 and 20 GPa, with most planetesimals equilibrating at between 5 and 30 GPa. Thus, a simulation characterised by deep magma oceans (because of the occurrence of very energetic giant impacts) has a small PEF and a simulation with more shallow magma oceans has a larger PEF.

We remind that the value of PEF should, in reality, be expected to vary for each planetesimal as they collide with a crystalising magma ocean. The value of PEF that we fit for represents an average of these equilibration conditions. A low value of PEF means that most planetesimal metal equilibrates progressively at the base of a shallow magma ocean towards the end of its crystallisation. A high value of PEF indicates that this metal equilibrates with a newly formed magma ocean. In reality, magma ocean solidification likely speeds up exponentially as it gets shallower meaning that higher values of PEF are expected. Thus, perhaps the simulations requiring the lowest values of PEF simply signal that the giant impacts are too energetic and the magma oceans are too deep with respect to the values characterising the real Earth history.

The compositions shown in \autoref{fig:PieCharts} are accompanied by the core mass fraction for the Earth analogues produced by the individual simulations. These are all slightly lower than the estimated core mass value of 32.59\% \citep{Valencia2006} though the uncertainty on this measurement combined with the uncertainties on our model likely bring the cores of the Earth-analogues to be consistent with that of the Earth.

We now focus on 7 simulations which fit at least 11 of the 12 oxides shown in \autoref{fig:Elements}: Sim\_74 Hot, Sim\_22 Cold, Sim\_74 Cold, Sim\_83 Cold, Sim\_33 Hot, Sim\_69 Hot and Sim\_83 Hot. 

Among their compositional pies (top seven charts lines in \autoref{fig:PieCharts}) we see all three of the above-mentioned groups, with the most common being group 3. The enrichment of the unsampled regions of these simulations span the large range of EF values and with the same trends shown in \autoref{fig:Enrichment_Changes}. 6 of the 7 simulations have a PEF that results in most planetesimal material equilibrating at pressures between 5 and 30 GPa, in line with the rest of the simulations. All 7 Earth analogues accreted from a mixture of all four compositional regions though many accrete very low quantities of enstatite or CI material. The CI material is expected to be low, accounting for 5\% of the Earth-material (from isotopic considerations \cite{Burkhardt2021}). In 4 out of 7 of these simulations, this material makes up less than 2\% of the final Earth composition. An addition of 1-2 CI planetesimals would bring this up to 5\% without substantially altering the final Earth composition. The low fraction of enstatite chondrites accreted to some Earth analogues is consistent with the well-documented paradox that, although the BSE is isotopically enstatite-like, it is very different for an elemental point of view \citep{Fitoussi2012,Dauphas2017,Morbidelli2020}. This could be explained if our refractory enriched, volatile depleted material, which is unsampled in the meteoritic record, had enstatite-like isotopic anomalies. This further implies that most of this material was distinct from the angrite parent body (as the latter has isotopic properties distinct from enstatite chondrites) and is thus truly unsampled in the meteorite collection.

We can now contemplate the most likely recipe of the three groups shown in \autoref{fig:PieCharts} in terms of Earth formation. It seems logical to favour group three because this contains many of the very good fits and also because it seems unlikely that the Earth would avoid accreting any enstatite chondrite-like compositions if they were present during formation. 

Sim\_74 Hot manages to fit all 12 components making this our overall best-fit simulation. Furthermore, the recipe for this Earth analogue requires 5.16\% carbonaceous material, a value very close to the 5\% estimate for the Earth from isotopic constraints. Additionally, the quantities of each meteorite bulk composition that makes up this Earth analogue seem viable to explain the volatile budget of the BSE and its characteristic depletion pattern (Sossi, private communication, March 2024). Further studies incorporating data from nucleosynthetic anomalies and radioactive chronometers into our model will enable us to truly see if this is a viable Earth formation scenario.

\section{Conclusions}
This study investigated the possibility of reproducing the chemical composition of the BSE based on a model in which the terrestrial planets accreted material mainly from a narrow local ring of planetesimals. In particular, we considered the Earth-formation simulations of \citet{Nesvorny2021} and adapted the work of \citet{Rubie2015} and \citet{Dale2023} to produce Earth-analogue mantles with elemental compositions similar to the BSE from the metal-silicate equilibration events associated with the sequences of accretional collisions within these simulations. We successfully fit the BSE assuming that the ring of planetesimals and embryos, with its low-density extension into the asteroid belt, comprises four compositional regions, in order of increasing distance from the Sun: a highly reduced component that is refractory enriched and unsampled in the meteoritic record, enstatite chondrites, ordinary chondrites and CI chondrites.  The compositional structure of the ring was allowed to vary according to four free parameters; three of which define the distances between the compositional regions and one which defines the enrichment of elements more refractory than Si relative to CI composition for the unsampled region. A fifth free parameter, constraining the average pressure and temperature of planetesimal metal-silicate equilibration events, was also varied throughout the model. 

Of 48 test simulations, which all produced a planet with around 1 Earth mass and which experienced a late giant impact, 17 resulted in an Earth analogue with a mantle that was roughly consistent with the composition of the BSE. The relative contributions of the four considered sources of material depend on the hierarchical sequence of collisions that creates the Earth analogue while the enrichment of the refractory enriched region strongly depends on its fractional abundance. None of the fits we obtained is perfect, showing the difficulty of reproducing the Earth within a given dynamical model, even with several free parameters. Nevertheless, we consider them instructive as to the actual formation conditions of the Earth. 

We form 7 Earth analogues with mantle compositions that more closely match the BSE. These 7 best fits create Earth from around 70\% of the refractory enriched component, further affirming the need for a material reservoir unsampled in the meteoritical collection for Earth formation. The PEF in these most successful simulations varies greatly suggesting the final elemental concentrations in the mantle are particularly sensitive to the equilibration conditions of planetesimal metal. One simulation fits all 12 considered chemical components of the BSE very well. It also matches other constraints on Earth formation such as the volatile budget of the BSE and the amount of carbonaceous material required to be accreted by Earth from isotopic constraints, at least at the qualitative level. The viability of this and other simulations should be further assessed by incorporating nucleosynthetic anomalies and radioactive chronometers into our models. For example, the timescales of accretion of the best-fit models vary greatly, with the ages of the final, moon-forming, giant impact ranging from around 10 to 125 Myr  (\autoref{fig:Earth_Growth}). Thus, modelling the $^{182}$W anomaly of Earth's mantle and comparing the age of the final giant impact with independently-determined ages of the moon, will result in additional tight constraints on the viability of accretion simulations \citep{Rubie2025}.

\section{Bibliography}

\bibliographystyle{cas-model2-names}

\bibliography{references}

\begin{thebibliography}{37}
\expandafter\ifx\csname natexlab\endcsname\relax\def\natexlab#1{#1}\fi
\providecommand{\url}[1]{\texttt{#1}}
\providecommand{\href}[2]{#2}
\providecommand{\path}[1]{#1}
\providecommand{\DOIprefix}{doi:}
\providecommand{\ArXivprefix}{arXiv:}
\providecommand{\URLprefix}{URL: }
\providecommand{\Pubmedprefix}{pmid:}
\providecommand{\doi}[1]{\href{http://dx.doi.org/#1}{\path{#1}}}
\providecommand{\Pubmed}[1]{\href{pmid:#1}{\path{#1}}}
\providecommand{\bibinfo}[2]{#2}
\ifx\xfnm\relax \def\xfnm[#1]{\unskip,\space#1}\fi
\bibitem[{{Alexander}(2019a)}]{Alexander2019A}
\bibinfo{author}{{Alexander}, C.M.O.}, \bibinfo{year}{2019}a.
\newblock \bibinfo{title}{{Quantitative models for the elemental and isotopic fractionations in chondrites: The carbonaceous chondrites}}.
\newblock \bibinfo{journal}{\gca} \bibinfo{volume}{254}, \bibinfo{pages}{277--309}.
\newblock \DOIprefix\doi{10.1016/j.gca.2019.02.008}.
\bibitem[{{Alexander}(2019b)}]{Alexander2019B}
\bibinfo{author}{{Alexander}, C.M.O.}, \bibinfo{year}{2019}b.
\newblock \bibinfo{title}{{Quantitative models for the elemental and isotopic fractionations in the chondrites: The non-carbonaceous chondrites}}.
\newblock \bibinfo{journal}{\gca} \bibinfo{volume}{254}, \bibinfo{pages}{246--276}.
\newblock \DOIprefix\doi{10.1016/j.gca.2019.01.026}.
\bibitem[{{Blanchard} et~al.(2022){Blanchard}, {Rubie}, {Jennings}, {Franchi}, {Zhao}, {Petitgirard}, {Miyajima}, {Jacobson} and {Morbidelli}}]{Blanchard2022}
\bibinfo{author}{{Blanchard}, I.}, \bibinfo{author}{{Rubie}, D.C.}, \bibinfo{author}{{Jennings}, E.S.}, \bibinfo{author}{{Franchi}, I.A.}, \bibinfo{author}{{Zhao}, X.}, \bibinfo{author}{{Petitgirard}, S.}, \bibinfo{author}{{Miyajima}, N.}, \bibinfo{author}{{Jacobson}, S.A.}, \bibinfo{author}{{Morbidelli}, A.}, \bibinfo{year}{2022}.
\newblock \bibinfo{title}{{The metal-silicate partitioning of carbon during Earth's accretion and its distribution in the early solar system}}.
\newblock \bibinfo{journal}{Earth and Planetary Science Letters} \bibinfo{volume}{580}, \bibinfo{pages}{117374}.
\newblock \DOIprefix\doi{10.1016/j.epsl.2022.117374}, \href{http://arxiv.org/abs/2202.06809}{\tt arXiv:2202.06809}.
\bibitem[{{Burkhardt} et~al.(2021){Burkhardt}, {Spitzer}, {Morbidelli}, {Budde}, {Render}, {Kruijer} and {Kleine}}]{Burkhardt2021}
\bibinfo{author}{{Burkhardt}, C.}, \bibinfo{author}{{Spitzer}, F.}, \bibinfo{author}{{Morbidelli}, A.}, \bibinfo{author}{{Budde}, G.}, \bibinfo{author}{{Render}, J.H.}, \bibinfo{author}{{Kruijer}, T.S.}, \bibinfo{author}{{Kleine}, T.}, \bibinfo{year}{2021}.
\newblock \bibinfo{title}{{Terrestrial planet formation from lost inner solar system material}}.
\newblock \bibinfo{journal}{Science Advances} \bibinfo{volume}{7}, \bibinfo{pages}{eabj7601}.
\newblock \DOIprefix\doi{10.1126/sciadv.abj7601}.
\bibitem[{{Chambers} and {Wetherill}(1998)}]{ChamberandWetherill1998}
\bibinfo{author}{{Chambers}, J.E.}, \bibinfo{author}{{Wetherill}, G.W.}, \bibinfo{year}{1998}.
\newblock \bibinfo{title}{{Making the Terrestrial Planets: N-Body Integrations of Planetary Embryos in Three Dimensions}}.
\newblock \bibinfo{journal}{\icarus} \bibinfo{volume}{136}, \bibinfo{pages}{304--327}.
\newblock \DOIprefix\doi{10.1006/icar.1998.6007}.
\bibitem[{{Chou}(1978)}]{Chou1978}
\bibinfo{author}{{Chou}, C.L.}, \bibinfo{year}{1978}.
\newblock \bibinfo{title}{{Abundances of Noble Metals in the Earth's Upper Mantle: Evidence for Late Heavy Bombardment After Core Formation}}.
\newblock \bibinfo{journal}{Meteoritics} \bibinfo{volume}{13}, \bibinfo{pages}{407}.
\bibitem[{{Dale} et~al.(2023){Dale}, {Rubie}, {Nakajima}, {Jacobson}, {Nathan}, {Golabek}, {Cambioni} and {Morbidelli}}]{Dale2023}
\bibinfo{author}{{Dale}, K.I.}, \bibinfo{author}{{Rubie}, D.C.}, \bibinfo{author}{{Nakajima}, M.}, \bibinfo{author}{{Jacobson}, S.}, \bibinfo{author}{{Nathan}, G.}, \bibinfo{author}{{Golabek}, G.J.}, \bibinfo{author}{{Cambioni}, S.}, \bibinfo{author}{{Morbidelli}, A.}, \bibinfo{year}{2023}.
\newblock \bibinfo{title}{{An improved model of metal/silicate differentiation during Earth's accretion}}.
\newblock \bibinfo{journal}{\icarus} \bibinfo{volume}{406}, \bibinfo{pages}{115739}.
\newblock \DOIprefix\doi{10.1016/j.icarus.2023.115739}, \href{http://arxiv.org/abs/2308.04476}{\tt arXiv:2308.04476}.
\bibitem[{{Dauphas}(2017)}]{Dauphas2017}
\bibinfo{author}{{Dauphas}, N.}, \bibinfo{year}{2017}.
\newblock \bibinfo{title}{{The isotopic nature of the Earth{\textquoteright}s accreting material through time}}.
\newblock \bibinfo{journal}{\nat} \bibinfo{volume}{541}, \bibinfo{pages}{521--524}.
\newblock \DOIprefix\doi{10.1038/nature20830}.
\bibitem[{{Dauphas} et~al.(2015){Dauphas}, {Poitrasson}, {Burkhardt}, {Kobayashi} and {Kurosawa}}]{Dauphas2015}
\bibinfo{author}{{Dauphas}, N.}, \bibinfo{author}{{Poitrasson}, F.}, \bibinfo{author}{{Burkhardt}, C.}, \bibinfo{author}{{Kobayashi}, H.}, \bibinfo{author}{{Kurosawa}, K.}, \bibinfo{year}{2015}.
\newblock \bibinfo{title}{{Planetary and meteoritic Mg/Si and {\ensuremath{\delta}}$^{30}$ Si variations inherited from solar nebula chemistry}}.
\newblock \bibinfo{journal}{Earth and Planetary Science Letters} \bibinfo{volume}{427}, \bibinfo{pages}{236--248}.
\newblock \DOIprefix\doi{10.1016/j.epsl.2015.07.008}, \href{http://arxiv.org/abs/1507.02922}{\tt arXiv:1507.02922}.
\bibitem[{{Deguen} et~al.(2014){Deguen}, {Landeau} and {Olson}}]{Deguen2014}
\bibinfo{author}{{Deguen}, R.}, \bibinfo{author}{{Landeau}, M.}, \bibinfo{author}{{Olson}, P.}, \bibinfo{year}{2014}.
\newblock \bibinfo{title}{{Turbulent metal-silicate mixing, fragmentation, and equilibration in magma oceans}}.
\newblock \bibinfo{journal}{Earth and Planetary Science Letters} \bibinfo{volume}{391}, \bibinfo{pages}{274--287}.
\newblock \DOIprefix\doi{10.1016/j.epsl.2014.02.007}, \href{http://arxiv.org/abs/1402.1666}{\tt arXiv:1402.1666}.
\bibitem[{{Elkins-Tanton}(2008)}]{ElkinsTanton2008}
\bibinfo{author}{{Elkins-Tanton}, L.T.}, \bibinfo{year}{2008}.
\newblock \bibinfo{title}{{Linked magma ocean solidification and atmospheric growth for Earth and Mars}}.
\newblock \bibinfo{journal}{Earth and Planetary Science Letters} \bibinfo{volume}{271}, \bibinfo{pages}{181--191}.
\newblock \DOIprefix\doi{10.1016/j.epsl.2008.03.062}.
\bibitem[{{Fischer-G{\"o}dde} and {Becker}(2012)}]{Fischer_Goode2012}
\bibinfo{author}{{Fischer-G{\"o}dde}, M.}, \bibinfo{author}{{Becker}, H.}, \bibinfo{year}{2012}.
\newblock \bibinfo{title}{{Osmium isotope and highly siderophile element constraints on ages and nature of meteoritic components in ancient lunar impact rocks}}.
\newblock \bibinfo{journal}{\gca} \bibinfo{volume}{77}, \bibinfo{pages}{135--156}.
\newblock \DOIprefix\doi{10.1016/j.gca.2011.11.014}.
\bibitem[{{Fitoussi} and {Bourdon}(2012)}]{Fitoussi2012}
\bibinfo{author}{{Fitoussi}, C.}, \bibinfo{author}{{Bourdon}, B.}, \bibinfo{year}{2012}.
\newblock \bibinfo{title}{{Silicon Isotope Evidence Against an Enstatite Chondrite Earth}}.
\newblock \bibinfo{journal}{Science} \bibinfo{volume}{335}, \bibinfo{pages}{1477}.
\newblock \DOIprefix\doi{10.1126/science.1219509}.
\bibitem[{{Greber} et~al.(2015){Greber}, {Puchtel}, {N{\"a}gler} and {Mezger}}]{Greber2015}
\bibinfo{author}{{Greber}, N.D.}, \bibinfo{author}{{Puchtel}, I.S.}, \bibinfo{author}{{N{\"a}gler}, T.F.}, \bibinfo{author}{{Mezger}, K.}, \bibinfo{year}{2015}.
\newblock \bibinfo{title}{{Komatiites constrain molybdenum isotope composition of the Earth's mantle}}.
\newblock \bibinfo{journal}{Earth and Planetary Science Letters} \bibinfo{volume}{421}, \bibinfo{pages}{129--138}.
\newblock \DOIprefix\doi{10.1016/j.epsl.2015.03.051}.
\bibitem[{{Hansen}(2009)}]{Hansen2009}
\bibinfo{author}{{Hansen}, B.M.S.}, \bibinfo{year}{2009}.
\newblock \bibinfo{title}{{Formation of the Terrestrial Planets from a Narrow Annulus}}.
\newblock \bibinfo{journal}{\apj} \bibinfo{volume}{703}, \bibinfo{pages}{1131--1140}.
\newblock \DOIprefix\doi{10.1088/0004-637X/703/1/1131}, \href{http://arxiv.org/abs/0908.0743}{\tt arXiv:0908.0743}.
\bibitem[{{Hutchison} et~al.(1987){Hutchison}, {Alexander} and {Barber}}]{Hutchison1987}
\bibinfo{author}{{Hutchison}, R.}, \bibinfo{author}{{Alexander}, C.M.O.}, \bibinfo{author}{{Barber}, D.J.}, \bibinfo{year}{1987}.
\newblock \bibinfo{title}{{The Semarkona meteorite: First recorded occurrence of smectite in an ordinary chondrite, and its implications}}.
\newblock \bibinfo{journal}{\gca} \bibinfo{volume}{51}, \bibinfo{pages}{1875--1882}.
\newblock \DOIprefix\doi{10.1016/0016-7037(87)90178-5}.
\bibitem[{{Izidoro} et~al.(2022){Izidoro}, {Dasgupta}, {Raymond}, {Deienno}, {Bitsch} and {Isella}}]{Izidoro2022}
\bibinfo{author}{{Izidoro}, A.}, \bibinfo{author}{{Dasgupta}, R.}, \bibinfo{author}{{Raymond}, S.N.}, \bibinfo{author}{{Deienno}, R.}, \bibinfo{author}{{Bitsch}, B.}, \bibinfo{author}{{Isella}, A.}, \bibinfo{year}{2022}.
\newblock \bibinfo{title}{{Planetesimal rings as the cause of the Solar System's planetary architecture}}.
\newblock \bibinfo{journal}{Nature Astronomy} \bibinfo{volume}{6}, \bibinfo{pages}{357--366}.
\newblock \DOIprefix\doi{10.1038/s41550-021-01557-z}, \href{http://arxiv.org/abs/2112.15558}{\tt arXiv:2112.15558}.
\bibitem[{{Jacobson} et~al.(2014){Jacobson}, {Morbidelli}, {Raymond}, {O'Brien}, {Walsh} and {Rubie}}]{Jacobson2014}
\bibinfo{author}{{Jacobson}, S.A.}, \bibinfo{author}{{Morbidelli}, A.}, \bibinfo{author}{{Raymond}, S.N.}, \bibinfo{author}{{O'Brien}, D.P.}, \bibinfo{author}{{Walsh}, K.J.}, \bibinfo{author}{{Rubie}, D.C.}, \bibinfo{year}{2014}.
\newblock \bibinfo{title}{{Highly siderophile elements in Earth's mantle as a clock for the Moon-forming impact}}.
\newblock \bibinfo{journal}{\nat} \bibinfo{volume}{508}, \bibinfo{pages}{84--87}.
\newblock \DOIprefix\doi{10.1038/nature13172}, \href{http://arxiv.org/abs/1504.01421}{\tt arXiv:1504.01421}.
\bibitem[{{Jennings} et~al.(2021){Jennings}, {Jacobson}, {Rubie}, {Nakajima}, {Vogel}, {Rose-Weston} and {Frost}}]{Jennings2021}
\bibinfo{author}{{Jennings}, E.S.}, \bibinfo{author}{{Jacobson}, S.A.}, \bibinfo{author}{{Rubie}, D.C.}, \bibinfo{author}{{Nakajima}, Y.}, \bibinfo{author}{{Vogel}, A.K.}, \bibinfo{author}{{Rose-Weston}, L.A.}, \bibinfo{author}{{Frost}, D.J.}, \bibinfo{year}{2021}.
\newblock \bibinfo{title}{{Metal-silicate partitioning of W and Mo and the role of carbon in controlling their abundances in the bulk silicate earth}}.
\newblock \bibinfo{journal}{\gca} \bibinfo{volume}{293}, \bibinfo{pages}{40--69}.
\newblock \DOIprefix\doi{10.1016/j.gca.2020.09.035}, \href{http://arxiv.org/abs/2210.14028}{\tt arXiv:2210.14028}.
\bibitem[{{Keil}(1968)}]{Keil1968}
\bibinfo{author}{{Keil}, K.}, \bibinfo{year}{1968}.
\newblock \bibinfo{title}{{Mineralogical and chemical relationships among enstatite chondrites}}.
\newblock \bibinfo{journal}{\jgr} \bibinfo{volume}{73}, \bibinfo{pages}{6945--6976}.
\newblock \DOIprefix\doi{10.1029/JB073i022p06945}.
\bibitem[{{Krot} et~al.(2014){Krot}, {Keil}, {Scott}, {Goodrich} and {Weisberg}}]{Krot2014}
\bibinfo{author}{{Krot}, A.N.}, \bibinfo{author}{{Keil}, K.}, \bibinfo{author}{{Scott}, E.R.D.}, \bibinfo{author}{{Goodrich}, C.A.}, \bibinfo{author}{{Weisberg}, M.K.}, \bibinfo{year}{2014}.
\newblock \bibinfo{title}{{Classification of Meteorites and Their Genetic Relationships}}, in: \bibinfo{editor}{{Davis}, A.M.} (Ed.), \bibinfo{booktitle}{Meteorites and Cosmochemical Processes}. volume~\bibinfo{volume}{1}, pp. \bibinfo{pages}{1--63}.
\bibitem[{{Lehner} et~al.(2014){Lehner}, {McDonough} and {N{\'e}Meth}}]{Lehner2014}
\bibinfo{author}{{Lehner}, S.W.}, \bibinfo{author}{{McDonough}, W.F.}, \bibinfo{author}{{N{\'e}Meth}, P.}, \bibinfo{year}{2014}.
\newblock \bibinfo{title}{{EH3 matrix mineralogy with major and trace element composition compared to chondrules}}.
\newblock \bibinfo{journal}{\maps} \bibinfo{volume}{49}, \bibinfo{pages}{2219--2240}.
\newblock \DOIprefix\doi{10.1111/maps.12391}.
\bibitem[{{Martin} and {Nokes}(1988)}]{MartinNokes1988}
\bibinfo{author}{{Martin}, D.}, \bibinfo{author}{{Nokes}, R.}, \bibinfo{year}{1988}.
\newblock \bibinfo{title}{{Crystal settling in a vigorously converting magma chamber}}.
\newblock \bibinfo{journal}{\nat} \bibinfo{volume}{332}, \bibinfo{pages}{534--536}.
\newblock \DOIprefix\doi{10.1038/332534a0}.
\bibitem[{{Morbidelli} et~al.(2020){Morbidelli}, {Libourel}, {Palme}, {Jacobson} and {Rubie}}]{Morbidelli2020}
\bibinfo{author}{{Morbidelli}, A.}, \bibinfo{author}{{Libourel}, G.}, \bibinfo{author}{{Palme}, H.}, \bibinfo{author}{{Jacobson}, S.A.}, \bibinfo{author}{{Rubie}, D.C.}, \bibinfo{year}{2020}.
\newblock \bibinfo{title}{{Subsolar Al/Si and Mg/Si ratios of non-carbonaceous chondrites reveal planetesimal formation during early condensation in the protoplanetary disk}}.
\newblock \bibinfo{journal}{Earth and Planetary Science Letters} \bibinfo{volume}{538}, \bibinfo{pages}{116220}.
\newblock \DOIprefix\doi{10.1016/j.epsl.2020.116220}, \href{http://arxiv.org/abs/2003.05486}{\tt arXiv:2003.05486}.
\bibitem[{{Nakajima} et~al.(2021){Nakajima}, {Golabek}, {W{\"u}nnemann}, {Rubie}, {Burger}, {Melosh}, {Jacobson}, {Manske} and {Hull}}]{Nakajima2021}
\bibinfo{author}{{Nakajima}, M.}, \bibinfo{author}{{Golabek}, G.J.}, \bibinfo{author}{{W{\"u}nnemann}, K.}, \bibinfo{author}{{Rubie}, D.C.}, \bibinfo{author}{{Burger}, C.}, \bibinfo{author}{{Melosh}, H.J.}, \bibinfo{author}{{Jacobson}, S.A.}, \bibinfo{author}{{Manske}, L.}, \bibinfo{author}{{Hull}, S.D.}, \bibinfo{year}{2021}.
\newblock \bibinfo{title}{{Scaling laws for the geometry of an impact-induced magma ocean}}.
\newblock \bibinfo{journal}{Earth and Planetary Science Letters} \bibinfo{volume}{568}, \bibinfo{pages}{116983}.
\newblock \DOIprefix\doi{10.1016/j.epsl.2021.116983}, \href{http://arxiv.org/abs/2004.04269}{\tt arXiv:2004.04269}.
\bibitem[{{Nesvorn{\'y}} et~al.(2021){Nesvorn{\'y}}, {Roig} and {Deienno}}]{Nesvorny2021}
\bibinfo{author}{{Nesvorn{\'y}}, D.}, \bibinfo{author}{{Roig}, F.V.}, \bibinfo{author}{{Deienno}, R.}, \bibinfo{year}{2021}.
\newblock \bibinfo{title}{{The Role of Early Giant-planet Instability in Terrestrial Planet Formation}}.
\newblock \bibinfo{journal}{\aj} \bibinfo{volume}{161}, \bibinfo{pages}{50}.
\newblock \DOIprefix\doi{10.3847/1538-3881/abc8ef}, \href{http://arxiv.org/abs/2012.02323}{\tt arXiv:2012.02323}.
\bibitem[{{Palme} and {O'Neill}(2003)}]{Palme&ONeill2003}
\bibinfo{author}{{Palme}, H.}, \bibinfo{author}{{O'Neill}, H.S.C.}, \bibinfo{year}{2003}.
\newblock \bibinfo{title}{{Cosmochemical Estimates of Mantle Composition}}.
\newblock \bibinfo{journal}{Treatise on Geochemistry} \bibinfo{volume}{2}, \bibinfo{pages}{568}.
\newblock \DOIprefix\doi{10.1016/B0-08-043751-6/02177-0}.
\bibitem[{{Piani} et~al.(2015){Piani}, {Robert} and {Remusat}}]{Piani2015}
\bibinfo{author}{{Piani}, L.}, \bibinfo{author}{{Robert}, F.}, \bibinfo{author}{{Remusat}, L.}, \bibinfo{year}{2015}.
\newblock \bibinfo{title}{{Micron-scale D/H heterogeneity in chondrite matrices: A signature of the pristine solar system water?}}
\newblock \bibinfo{journal}{Earth and Planetary Science Letters} \bibinfo{volume}{415}, \bibinfo{pages}{154--164}.
\newblock \DOIprefix\doi{10.1016/j.epsl.2015.01.039}, \href{http://arxiv.org/abs/1502.01067}{\tt arXiv:1502.01067}.
\bibitem[{{Rubie} et~al.(2025){Rubie}, {Dale}, {Nathan}, {Nakajima}, {Jennings}, {Golabek}, {Jacobson} and {Morbidelli}}]{Rubie2025}
\bibinfo{author}{{Rubie}, D.C.}, \bibinfo{author}{{Dale}, K.I.}, \bibinfo{author}{{Nathan}, G.}, \bibinfo{author}{{Nakajima}, M.}, \bibinfo{author}{{Jennings}, E.S.}, \bibinfo{author}{{Golabek}, G.J.}, \bibinfo{author}{{Jacobson}, S.A.}, \bibinfo{author}{{Morbidelli}, A.}, \bibinfo{year}{2025}.
\newblock \bibinfo{title}{{Tungsten isotope evolution during Earth's formation and new constraints on the viability of accretion simulations}}.
\newblock \bibinfo{journal}{Earth and Planetary Science Letters} \bibinfo{volume}{651}, \bibinfo{pages}{119139}.
\newblock \DOIprefix\doi{10.1016/j.epsl.2024.119139}, \href{http://arxiv.org/abs/2411.17681}{\tt arXiv:2411.17681}.
\bibitem[{{Rubie} et~al.(2011){Rubie}, {Frost}, {Mann}, {Asahara}, {Nimmo}, {Tsuno}, {Kegler}, {Holzheid} and {Palme}}]{Rubie2011}
\bibinfo{author}{{Rubie}, D.C.}, \bibinfo{author}{{Frost}, D.J.}, \bibinfo{author}{{Mann}, U.}, \bibinfo{author}{{Asahara}, Y.}, \bibinfo{author}{{Nimmo}, F.}, \bibinfo{author}{{Tsuno}, K.}, \bibinfo{author}{{Kegler}, P.}, \bibinfo{author}{{Holzheid}, A.}, \bibinfo{author}{{Palme}, H.}, \bibinfo{year}{2011}.
\newblock \bibinfo{title}{{Heterogeneous accretion, composition and core-mantle differentiation of the Earth}}.
\newblock \bibinfo{journal}{Earth and Planetary Science Letters} \bibinfo{volume}{301}, \bibinfo{pages}{31--42}.
\newblock \DOIprefix\doi{10.1016/j.epsl.2010.11.030}.
\bibitem[{{Rubie} et~al.(2015){Rubie}, {Jacobson}, {Morbidelli}, {O'Brien}, {Young}, {de Vries}, {Nimmo}, {Palme} and {Frost}}]{Rubie2015}
\bibinfo{author}{{Rubie}, D.C.}, \bibinfo{author}{{Jacobson}, S.A.}, \bibinfo{author}{{Morbidelli}, A.}, \bibinfo{author}{{O'Brien}, D.P.}, \bibinfo{author}{{Young}, E.D.}, \bibinfo{author}{{de Vries}, J.}, \bibinfo{author}{{Nimmo}, F.}, \bibinfo{author}{{Palme}, H.}, \bibinfo{author}{{Frost}, D.J.}, \bibinfo{year}{2015}.
\newblock \bibinfo{title}{{Accretion and differentiation of the terrestrial planets with implications for the compositions of early-formed Solar System bodies and accretion of water}}.
\newblock \bibinfo{journal}{\icarus} \bibinfo{volume}{248}, \bibinfo{pages}{89--108}.
\newblock \DOIprefix\doi{10.1016/j.icarus.2014.10.015}, \href{http://arxiv.org/abs/1410.3509}{\tt arXiv:1410.3509}.
\bibitem[{{Rubie} et~al.(2003){Rubie}, {Melosh}, {Reid}, {Liebske} and {Righter}}]{Rubie2003}
\bibinfo{author}{{Rubie}, D.C.}, \bibinfo{author}{{Melosh}, H.J.}, \bibinfo{author}{{Reid}, J.E.}, \bibinfo{author}{{Liebske}, C.}, \bibinfo{author}{{Righter}, K.}, \bibinfo{year}{2003}.
\newblock \bibinfo{title}{{Mechanisms of metal-silicate equilibration in the terrestrial magma ocean}}.
\newblock \bibinfo{journal}{Earth and Planetary Science Letters} \bibinfo{volume}{205}, \bibinfo{pages}{239--255}.
\newblock \DOIprefix\doi{10.1016/S0012-821X(02)01044-0}.
\bibitem[{{Tissot} et~al.(2022){Tissot}, {Collinet}, {Namur} and {Grove}}]{Tissot2022}
\bibinfo{author}{{Tissot}, F.L.H.}, \bibinfo{author}{{Collinet}, M.}, \bibinfo{author}{{Namur}, O.}, \bibinfo{author}{{Grove}, T.L.}, \bibinfo{year}{2022}.
\newblock \bibinfo{title}{{The case for the angrite parent body as the archetypal first-generation planetesimal: Large, reduced and Mg-enriched}}.
\newblock \bibinfo{journal}{\gca} \bibinfo{volume}{338}, \bibinfo{pages}{278--301}.
\newblock \DOIprefix\doi{10.1016/j.gca.2022.09.031}.
\bibitem[{{Valencia} et~al.(2006){Valencia}, {O'Connell} and {Sasselov}}]{Valencia2006}
\bibinfo{author}{{Valencia}, D.}, \bibinfo{author}{{O'Connell}, R.J.}, \bibinfo{author}{{Sasselov}, D.}, \bibinfo{year}{2006}.
\newblock \bibinfo{title}{{Internal structure of massive terrestrial planets}}.
\newblock \bibinfo{journal}{\icarus} \bibinfo{volume}{181}, \bibinfo{pages}{545--554}.
\newblock \DOIprefix\doi{10.1016/j.icarus.2005.11.021}, \href{http://arxiv.org/abs/astro-ph/0511150}{\tt arXiv:astro-ph/0511150}.
\bibitem[{{Walsh} et~al.(2011){Walsh}, {Morbidelli}, {Raymond}, {O'Brien} and {Mandell}}]{Walsh2011}
\bibinfo{author}{{Walsh}, K.J.}, \bibinfo{author}{{Morbidelli}, A.}, \bibinfo{author}{{Raymond}, S.N.}, \bibinfo{author}{{O'Brien}, D.P.}, \bibinfo{author}{{Mandell}, A.M.}, \bibinfo{year}{2011}.
\newblock \bibinfo{title}{{A low mass for Mars from Jupiter's early gas-driven migration}}.
\newblock \bibinfo{journal}{\nat} \bibinfo{volume}{475}, \bibinfo{pages}{206--209}.
\newblock \DOIprefix\doi{10.1038/nature10201}, \href{http://arxiv.org/abs/1201.5177}{\tt arXiv:1201.5177}.
\bibitem[{{Woo} et~al.(2023){Woo}, {Morbidelli}, {Grimm}, {Stadel} and {Brasser}}]{Woo2023}
\bibinfo{author}{{Woo}, J.M.Y.}, \bibinfo{author}{{Morbidelli}, A.}, \bibinfo{author}{{Grimm}, S.L.}, \bibinfo{author}{{Stadel}, J.}, \bibinfo{author}{{Brasser}, R.}, \bibinfo{year}{2023}.
\newblock \bibinfo{title}{{Terrestrial planet formation from a ring}}.
\newblock \bibinfo{journal}{\icarus} \bibinfo{volume}{396}, \bibinfo{pages}{115497}.
\newblock \DOIprefix\doi{10.1016/j.icarus.2023.115497}, \href{http://arxiv.org/abs/2302.14100}{\tt arXiv:2302.14100}.
\bibitem[{{Woo} et~al.(2024){Woo}, {Nesvorn{\'y}}, {Scora} and {Morbidelli}}]{Woo2024}
\bibinfo{author}{{Woo}, J.M.Y.}, \bibinfo{author}{{Nesvorn{\'y}}, D.}, \bibinfo{author}{{Scora}, J.}, \bibinfo{author}{{Morbidelli}, A.}, \bibinfo{year}{2024}.
\newblock \bibinfo{title}{{Terrestrial planet formation from a ring: Long-term simulations accounting for the giant planet instability}}.
\newblock \bibinfo{journal}{\icarus} \bibinfo{volume}{417}, \bibinfo{pages}{116109}.
\newblock \DOIprefix\doi{10.1016/j.icarus.2024.116109}, \href{http://arxiv.org/abs/2404.17259}{\tt arXiv:2404.17259}.

\end{thebibliography}
\newpage
\section*{Supplementary Material}
\renewcommand{\thefigure}{S.\arabic{figure}} 
\setcounter{figure}{0} 
\subsection*{Supression of innermost material }
\begin{figure*}[htbp!]
    \centering
  \includegraphics[width=0.75\textwidth]{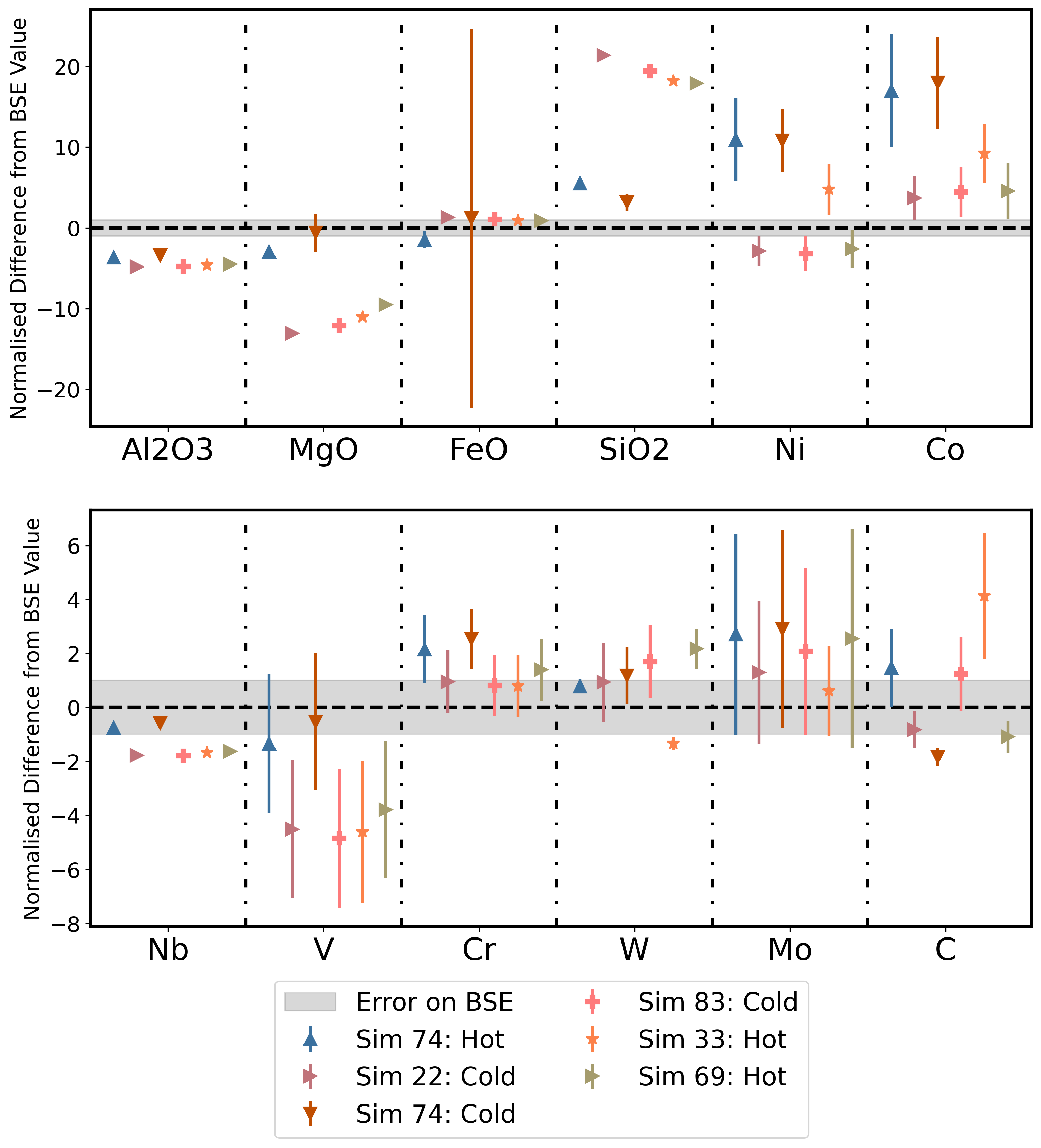} 
  \caption{The compositions of six Earth analogue mantles following a best-fit calculation with the material enriched in elements more refractory than Si relative to CI concentrations suppressed. All models were attempting to fit the BSE composition and all failed to form an Earth-analogue mantle with a composition similar to the BSE. While best-fits have not been calculated as described in the main text it should be noted that the chi-sq for all these simulations lies between 80 and 100, compared with 3-7 for the simulations discussed in the main text. The change in scale when compared to \autoref{fig:Elements} should also be noted.}
  \label{fig:NoEnrich}
\end{figure*}

\end{document}